\documentstyle[12pt]{article}

\font\blackboard=msbm10 at 12pt
\font\blackboards=msbm7
\font\blackboardss=msbm5
\newfam\black
\textfont\black=\blackboard
\scriptfont\black=\blackboards
\scriptscriptfont\black=\blackboardss

\newcommand{\junk}[1]{}

\newcommand{\ba}{\begin{array}}
\newcommand{\ea}{\end{array}}
\newcommand{\be}{\begin{equation}}
\newcommand{\ee}{\end{equation}}
\newcommand{\bea}{\begin{eqnarray}}
\newcommand{\eea}{\end{eqnarray}}
\newcommand{\beas}{\begin{eqnarray*}}
\newcommand{\eeas}{\end{eqnarray*}}

\def\laplace{{\kern1pt\vbox{\hrule height 1.2pt\hbox{\vrule width
1.2pt\hskip
  3pt\vbox{\vskip 6pt}\hskip 3pt\vrule width 0.6pt}\hrule height
  0.6pt}
  \kern1pt}}
\def\scriptlap{{\kern1pt\vbox{\hrule height 0.8pt\hbox{\vrule width
  0.8pt
  \hskip2pt\vbox{\vskip 4pt}\hskip 2pt\vrule width 0.4pt}\hrule height
  0.4pt}
  \kern1pt}}

\def\roughly#1{\raise.3ex\hbox{$#1$\kern-.75em\lower1ex\hbox{$\sim$}}}

\def\gym{g^2_{\scriptscriptstyle YM}}

\def\str{{\rm STr} \,}

\def\tr{{\rm Tr} \,}

\newcommand{\NP}{{\em Nucl.\ Phys.\ }}

\newcommand{\PL}{{\em Phys.\ Lett.\ }}
\newcommand{\PR}{{\em Phys.\ Rev.\ }}

\newcommand{\PRL}{{\em Phys.\ Rev.\ Lett.\ }}

\newcommand{\gone}[1]{}
\begin{document}
\pagestyle{plain}
\setcounter{page}{1}

\baselineskip16pt

\begin{titlepage}

\begin{flushright}
PUPT-1865\\
MIT-CTP-2866\\
hep-th/9905174
\end{flushright}
\vspace{8 mm}

\begin{center}

{\Large \bf   Absorption of dilaton partial waves by D3-branes\\}
%\vspace{3mm}

\end{center}

\vspace{7 mm}

\begin{center}

Igor Klebanov$^a$, Washington Taylor IV$^b$ and Mark Van Raamsdonk$^a$

\vspace{3mm}
${}^a${\small \sl Department of Physics} \\
{\small \sl Joseph Henry Laboratories} \\
{\small \sl Princeton University} \\
{\small \sl Princeton, New Jersey 08544, U.S.A.} \\
{\small \tt klebanov@puhep1.princeton.edu, mav@princeton.edu}

\vspace{3mm}
${}^b${\small \sl Center for Theoretical Physics} \\
{\small \sl MIT, Bldg. 6-306} \\
{\small \sl Cambridge, MA 02139, U.S.A.} \\
{\small \tt wati@mit.edu}\\
\end{center}

\vspace{8 mm}

\begin{abstract}
We calculate the leading term in the low-energy absorption cross section for
an arbitrary partial wave of the dilaton field by a 
stack of many coincident D3-branes. We find that it precisely
reproduces the semiclassical absorption cross section of a 
3-brane geometry, including all numerical factors.  The crucial ingredient in
making the correspondence is the identification of the precise
operators on the D3-brane world-volume which couple to the dilaton
field and all its derivatives.  The needed operators are related
through T-duality and the IIA/M-theory correspondence to the recently
determined M(atrix) theory expressions for multipole moments of the 11D
supercurrent.  These operators have a characteristic symmetrized trace
structure which plays a key combinatorial role in the analysis for the
higher partial waves.
The results presented here give  new evidence for an infinite
family of non-renormalization theorems which are believed to exist
for two-point functions in ${\cal N} = 4$ gauge theory in four
dimensions.
\end{abstract}

%\vspace{2cm}
\vspace{1cm}
\begin{flushleft}
May 1999
\end{flushleft}
\end{titlepage}
\newpage

%%%%%%%%%%%%%%%%%%%%%%%%%
\section{Introduction}

Black $p$-brane solutions of type II supergravity carrying
Ramond-Ramond (RR) charges have been known since the early 90's
\cite{Horowitz-strominger,Duff-Lu}.  The string frame metric and dilaton
backgrounds in such solutions may be expressed in the following simple
form: 
\be
\label{metric}
   ds^2 =
H^{-\frac{1}{2} }(r)
    \left[ - dt^2 + \sum_{i=1}^p (d x^i)^2 \right] +
H^{\frac{1}{2} }(r)
    \left[ dr^2 + r^2 d\Omega_{8-p}^2 \right] \ ,
\ee
$$ e^\Phi = H^{(3-p)/4}(r)\ ,
$$
where
$$   H(r)  = 1 + {R^{7-p} \over r^{7-p}} \ .  
$$
The importance of these solutions was not fully appreciated until
Polchinski realized that the Dirichlet $p$-brane is the elementary
object in string theory that couples to the $(p+1)$-form RR potential
\cite{Polchinski}.  This made it clear that the $p$-brane solutions of
\cite{Horowitz-strominger} describe the classical fields created by a
large number of coincident D$p$-branes. Since the low-energy
world-volume dynamics of $N$ parallel D$p$-branes is governed by
maximally supersymmetric $U(N)$ gauge theory \cite{Witten-bound}, this
suggests a relation between such a gauge theory in $ p + 1$ dimensions
and type II string theory in the background of the classical $p$-brane
solution.

Among early hints that this relationship between supersymmetric gauge
theory and string theory is exact was the calculation of the dilaton
absorption cross section by threebranes
\cite{Klebanov-absorption}.\footnote{This calculation was in turn
motivated by similar calculations in the D1-D5 system
\cite{Callan-Maldacena,dmw,Das-Mathur-2}. There such studies are more
difficult, however, due to the complexity of the world volume dynamics
of intersecting D-branes.}  The threebrane solution is of particular
interest because it is the only non-singular solution of the form
(\ref{metric}). 
Furthermore, the low-energy dynamics of coincident D3-branes
is described by ${\cal N}=4$ supersymmetric Yang-Mills theory, which
is an attractive theory because of its exact conformal
invariance. A related fact is that the dilaton background is constant,
so that the dilaton fluctuation satisfies the minimally coupled scalar
equation
\begin{equation}
\partial_\mu \left (\sqrt{-g}g^{\mu\nu}\partial_\nu \phi\right )=0\ .
\end{equation}
In \cite{Klebanov-absorption} this equation was solved for incident
s-waves of low-energy $\omega$.
The leading term in the absorption cross section  was calculated to be
\be
\label{three}
\sigma_{SUGRA}= {\pi^4\over 8}\omega^3 R^8 \ .  \ee 
This result was compared to a corresponding calculation in the SYM
theory, where the dilaton couples to the operator ${T_3\over 4}\tr
(F^2+\cdots)$ ($T_3$ is the D3-brane tension).  At weak coupling the
leading order absorption process is for the dilaton to turn into a
pair of gluons on the world-volume. The rate for this process was
calculated in the 3-brane gauge theory and was found to be  
\cite{Klebanov-absorption}
\be
\label{absorb} \sigma = {\kappa^2 \omega^3 N^2\over 32 \pi} \ , 
\ee
Remarkably, this is equal to (\ref{three}) after we take into account
the relation 
\be
\label{throatrel} R^4 = {\kappa\over 2\pi^{5/2}} N \ ,
\ee which can be found by equating the tensions of the black 3-brane and 
$N$
D3-branes \cite{gkp-3}.  This equality of the low-energy
cross sections raises the hope of an {\it exact} relation between SYM
theory and gravity.  There seems to be a puzzle, however, because the
gravitational calculation becomes reliable in the weak curvature limit
where $g_{{\rm YM}}^2 N\rightarrow \infty$ while the SYM calculation
was carried out to leading order in $g_{{\rm YM}}^2 N$.  In
\cite{Gubser-Klebanov} this puzzle was resolved by arguing that all
higher order corrections in the coupling vanish due to supersymmetric
non-renormalization theorems. (This theorem was made explicit for the
absorption cross section of gravitons calculated in \cite{gkt}, which
is related to the 2-point function of the stress-energy tensor of the
gauge theory.) Thus, the agreement of s-wave cross sections found in
\cite{Klebanov-absorption,gkt} is actually necessary if SYM theory and
gravity are exactly related. This agreement is one of the pieces of
evidence in favor of the exact AdS/CFT correspondence between the 
threebrane
throat and the ${\cal N}=4$ SYM theory formulated in
\cite{Maldacena-AdS,gkp-2,Witten-AdS1}.

An immediate question is whether this agreement persists for the
absorption of higher partial waves. In \cite{Klebanov-absorption} it
was suggested that the operator responsible for absorption of the
$l$th partial wave of the dilaton should be of the form 
\be
\label{Ops} {T_3\over 4 l!}\tr( F_{ab} F^{ab} X^{(i_1} \cdots
X^{i_l)})_{\rm Traceless} \ .  
\ee 
It is not hard to show that the
operator (\ref{Ops}) leads to a SYM cross section which scales in the
same way with respect to $N$ and $\omega$ as the cross section
computed from the semiclassical gravity theory
\cite{Klebanov-absorption}.  In \cite{gkt} an attempt was made to
compare the constant factors in these cross sections, but the results
seemed discouraging: the SYM answer seemed to grow with $l$ much
faster than the gravity answer.

In this paper we resolve this problem and show that the gravity and
the SYM cross sections are in {\it exact} agreement for all $l$.
This is the first example of such a match occurring for all partial
waves.  Higher partial wave absorption processes were considered for
the D1 + D5 system in \cite{Maldacena-strominger,Mathur,Gubser};
because the world-volume theory of the branes is not as well
understood in that case, however, it is not yet possible to make a
precise numerical comparison between the supergravity and D-brane
predictions for the absorption cross section.

In order to do an exact absorption calculation in the D3-brane gauge
theory we need to know the precise operators in the world-volume super
Yang-Mills theory which couple linearly to the bulk dilaton field and
its derivatives.  These operators will contain terms of the form
(\ref{Ops}), but this is not a complete description of the operators
we need.  There is an ordering ambiguity in (\ref{Ops}) when $N > 1$.
There are also additional fermion terms which must be considered for
$l > 0$.  Analogous operators to those we need were recently computed
in \cite{Mark-Wati-4}, where recent results on the M(atrix) theory form
of the supercurrent in DLCQ M-theory
\cite{Mark-Wati,Dan-Wati-2,Mark-Wati-3} were used to find the
operators in the world volume theory of a system of D0-branes coupling
to weak IIA background fields.  By T-dualizing the results of
\cite{Mark-Wati-4} in three directions, we can determine the desired
D3-brane operators and use them to precisely compute the absorption
cross sections.  Let us consider the $l=1$ partial wave as an example.
In \cite{gkt} it was claimed that the operator ${T_3\over 4 }\tr
F_{ab} F^{ab} X^i$ gives a SYM absorption cross section which agrees
with the classical result. We have found, however, that when ordering
effects are taken into consideration this term accounts for only $1/2$
of the classical absorption cross section for the leading terms in a
large $N$ expansion.  Luckily, the results in \cite{Mark-Wati-4}
indicate that there is another operator contributing at the same
order: 
\be 
\label{newop} 
{T_3\over 16}\tr (F_{jk} \bar\Theta
\Gamma^{[jki]} \Theta - F_{ab} \bar\Theta \Gamma^{[abi]} \Theta ) \ ,
\ee 
where $F, \Theta$ and the matrices $\Gamma$ are written in 10D
notation with $a, b \in\{0,1,2,3\}$ and $i, j, k \in\{4,\ldots, 9\}$. 
This operator accounts
for the other half of the classical cross section and restores the
agreement for the $l=1$ partial wave.

For $l>1$, a two-fermion operator of the form of
(\ref{newop}) must again be included in the
cross section calculation.  In addition, there are quartic terms in
the fermions which appear at $l = 2$.  Although the four-fermion terms
in the relevant D0-brane and Matrix theory operators have not been
calculated, it is possible to fix these terms by using supersymmetry
and our knowledge of the bosonic terms.  The operators we need are 
essentially the same ones that correspond to Kaluza-Klein modes of the 
dilaton in the correspondence between $AdS^5 \times S^5$ and 
${\cal N}=4$ SYM theory, and can be
found by acting with 4 supercharges on the superconformal chiral primary
fields 
\be
\label{primary} 
{\cal O}_{(l+2)}^{{\rm cp}}\sim \tr (X^{(i_1} \ldots X^{i_{l+2})
})_{\rm Traceless} \ .  
\ee 
Since we only act with 4 supersymmetry transformations, we
encounter at most 4 fermion fields.  From our knowledge of the bosonic
and two-fermion components of the operators, it is possible to find
the proper combination of supersymmetry operators which give the
unique four-fermion extension of the lower order components compatible
with supersymmetry.

A key feature of the operators found in the study of Matrix theory
supercurrents is the symmetrized trace structure which dictates that
all traces should be averaged over orderings of the $N \times N$
matrices $F^{\mu \nu}, X^i, \Theta$ and $D \Theta$.  It was suggested
some time ago by Tseytlin \cite{Tseytlin} that the symmetrized trace
is the correct way to extend the abelian Born-Infeld action
to a nonabelian theory.
While for the full nonabelian Born-Infeld action this remains a 
conjecture,
we emphasize that for the operators we are interested in here this
structure has been deduced from an explicit
calculation in the Matrix theory context.  As further evidence for
this structure,
in \cite{Mark-Wati-4} it was shown that the symmetrized trace
gives rise to nontrivial combinatorial factors which allow the
D0-brane action in weakly curved backgrounds to satisfy the geodesic
length condition suggested by Douglas in \cite{Douglas-curved}.
In the present paper we find that the symmetrized trace structure and
the correct counting of graphs according to 't Hooft's large $N$ limit
are crucial in achieving  exact agreement between the D-brane
absorption calculation and the semiclassical results for $l > 0$.

In Section 2 we review the semiclassical calculation of the higher
partial wave absorption cross sections originally found in \cite{gkt}.
The complete construction of the world-volume
operators in the D3-brane theory is
presented in Section 3.  In Section 4 we calculate the 2-point
functions of these operators to leading order in $g_{{\rm YM}}^2 N$
and convert these results into
absorption cross sections, finding exact agreement with the semiclassical
calculations for all $l$.  Since the semiclassical calculations are valid
for $g_{{\rm YM}}^2 N\rightarrow \infty$, this is evidence in favor of
non-renormalization theorems protecting the 2-point functions of all
operators constructed in section 3.  In Section  5 we present a
discussion of our results and conclude.

\section{Semiclassical absorption calculation}

In this section we review the semiclassical calculation of the
absorption cross section for an arbitrary partial wave of the dilaton
in the extremal 3-brane background.  The results of this calculation
were originally given in \cite{gkt}.

The semiclassical approach to computing the absorption cross section
for a field propagating in a black hole background geometry was
pioneered in the thesis of Unruh \cite{Unruh}.  In recent times, this
method has been used to study the absorption cross section for fields
in the 5D black hole geometry produced by a D1 + D5 system
\cite{dmw,Das-Mathur-2} and in the 7D black hole geometry produced by
multiple D3-branes \cite{Klebanov-absorption,gkt}.  The first step in
a calculation of this type is to determine the field equation for a
fixed partial wave of the field of interest.  This wave equation can
usually be solved approximately in certain regimes of the radial
parameter $r$.  These approximate solutions are then matched between
regions and a solution is chosen which satisfies the boundary
condition that there is no outgoing flux at the horizon.  The
absorption coefficient is then given by the ratio of the inward flux
at the horizon over the inward flux at $r = \infty$.  Application of
the standard optical theorem from quantum mechanics gives the
absorption cross section in terms of the absorption coefficient.  It
is often necessary to obtain approximate solutions in three distinct
regions of the parameter $r$.  A particularly nice feature of the
minimally coupled scalars in the 3-brane background is that only two
regions are necessary, which simplifies some aspects of the story.

We now outline the application of this method to the dilaton in the
3-brane background, following \cite{Klebanov-absorption,gkt}.
The wave equation for the $l$th partial wave of a dilaton mode with
frequency $\omega$ in the
background (\ref{metric}) with $p = 3$ is
\begin{equation}
\left[\frac{1}{ \rho^5}  \frac{d}{d \rho}  \rho^5 \frac{d}{d \rho} + 1 +
 \frac{\omega^4 R^4}{\rho^4}  -\frac{l (l + 4)}{ \rho^2}  \right]
\phi^{(l)} (\rho) = 0
\label{eq:l-wave}
\end{equation}
where $\rho = \omega r$.
The absorption process we are interested in thus corresponds to 
quantum mechanical tunneling through a centrifugal potential barrier
in the reduced
one-dimensional system.
In both the large $\rho$ ($\rho \gg (\omega R)^2$) and small $\rho$
($\rho \ll 1$) regimes, (\ref{eq:l-wave}) reduces to a Bessel equation.
The solution for large $\rho$ is
\begin{equation}
\phi^{(l)} (\rho) = A \rho^{-2} J_{l + 2} (\rho)
+B \rho^{-2}  N_{l + 2} (\rho)
\label{eq:large-r}
\end{equation}
where $A, B$ are undetermined constants.  The solution in the regime
$\rho \ll 1$
is
\begin{equation}
\phi^{(l)} (\rho) = i (\omega R)^4 \rho^{-2}
\left[ J_{l + 2} \left( \frac{\omega^2 R^2}{ \rho}  \right)
+ iN_{l + 2} \left( \frac{\omega^2 R^2}{ \rho}  \right)\right]
\label{eq:small-r}
\end{equation}
where the overall normalization has been fixed to an arbitrary
constant and the relative coefficients $J_{l + 2} + i N_{l + 2}$ are
fixed by the condition that the flux at $\rho \rightarrow 0$ describes
a purely incoming wave.  In the overlap region $(\omega R)^2 \ll \rho
\ll 1$ we can use the asymptotic forms for the Bessel function to find
from (\ref{eq:small-r})
\[
\phi^{(l)} (\rho) \sim \frac{2^{l + 2} (l + 1)!\rho^l}{ \pi (\omega
R)^{2l}}  + {\rm subleading}
\]
This determines the coefficients $A, B$ in (\ref{eq:large-r})
to be
\[
A = \frac{4^{l + 2} (l + 1)!(l + 2)!}{\pi (\omega R)^{2l}}, \;\;\;\;\;
B = 0
\]
The absorption coefficient is then given by the ratio of the incoming
flux at $\rho = 0$ over the incoming flux at $\rho = \infty$
\begin{equation}
P_l = \frac{(\omega R)^{4l + 8} \pi^2}{ 4^{2l + 3} [(l + 1)!]^2[(l + 
2)!]^2} 
\label{eq:absorption-coefficient}
\end{equation}
The optical theorem in 7 space-time dimensions relates the absorption
cross section $\sigma_s^l$ to the absorption coefficient through
\cite{Gubser}
\begin{equation}
\sigma^l_s = \frac{8 \pi^2}{3 \omega^5} (l + 1) (l + 2)^2 (l + 3) P_l.
\label{eq:optical}
\end{equation}
Combining this with (\ref{eq:absorption-coefficient}) we find that the
semiclassical result for the leading order contribution to the
total absorption cross section is
\begin{equation}
\sigma^l_s = \frac{\pi^4}{ 24}  \frac{(l + 3) (l + 1)}{ [(l + 1)!]^4}
\left( \frac{\omega R}{ 2} \right)^{4l} \omega^3 R^8.
\end{equation}
Replacing $R^4$ through (\ref{throatrel}) this can be rewritten as
\begin{equation}
\sigma^l_s = \frac{N^{l + 2} \kappa^{l + 2} \omega^{4l + 3} (l + 3)}{
3 \cdot 2^{5l + 5} \pi^{5l/2 + 1}l![(l + 1)!]^3}.
\label{eq:semiclassical-absorption}
\end{equation}

The semiclassical result (\ref{eq:semiclassical-absorption}) is the
leading term in the absorption cross section for an arbitrary partial
wave $l$; this is the result which we will reproduce from the D-brane
point of view in the remainder of the paper.  Although the method we
have just outlined gives the correct answer at leading order in
$\omega R$ for each partial wave $l$, there are subleading corrections
to this result which may also be of interest.  These subleading
corrections were determined by Gubser and Hashimoto in
\cite{Gubser-Hashimoto}.  In that paper, it is shown that the wave
equation (\ref{eq:l-wave}) is equivalent to Mathieu's modified
differential equation
\[
\left[ \frac{\partial}{ \partial z^2}  + 2q \cosh 2z-a \right] \psi
(z) = 0.
\]
The exact solution of this equation is known as a power series in $q =
\omega R$.  This exact solution is used in \cite{Gubser-Hashimoto} to
write the complete expansion for the absorption probability of the
$l$th partial wave
\begin{equation}
P_l = \frac{4 \pi^2}{[(l + 1)!]^2[(l + 2)!^2]} \left( \frac{\omega R}{
2}  \right)^{8 + 4l} \sum_{0 \leq k \leq n} b_{n, k} (\omega R)^{4n}
(\ln \omega \gamma R/2)^k
\label{eq:exact-l}
\end{equation}
where $b_{n, k}$ are computable coefficients with $b_{0, 0} = 1$ and
$\ln \gamma$ is Euler's constant.  In the final section of this paper
we briefly discuss the possibility of extending the results in this
paper to include some of these higher order corrections.

\section{Coupling of the dilaton to the world volume theory}

In this section we determine how the type IIB dilaton field couples to
the world-volume theory on the branes.

The world volume theory of $N$ D3 branes is the $D=4$, ${\cal N} = 4$  
supersymmetric Yang-Mills theory with gauge group $U(N)$. This theory 
may be obtained as the dimensional reduction of $D=10$ super Yang-Mills 
theory, and throughout this work, we will use $D=10$ language, writing 
operators in terms of 32 component Majorana-Weyl spinors and $32 \times 
32$ gamma matrices. From a four dimensional perspective, these gamma 
matrices contain not only the four $D=4$ gamma matrices, but also 
Clebsch-Gordon coefficients relating the ${\bf 6}$ representation of the 
R symmetry group $SU(4)$ (equivalent to the fundamental representation 
of the $SO(6)$ manifest in the $D=10$ language) to the representation 
${\bf 4} \oplus {\bf \bar{4}}$ carried by a fermion bilinear in the 
$D=4$ language.  

   The coupling of the type IIB dilaton field to the world volume
theory of $N$ D3 branes is in principle given by the non-abelian
Born-Infeld action which sums all planar string diagrams
describing interactions between the lightest string fields on the D-brane
world-volume and in the bulk.  The complete form of this action is not 
known,
although it has been proposed that the background independent bosonic 
terms are those obtained by T-duality from a 9-brane action obtained from 
the abelian version by symmetrizing all traces \cite{Tseytlin}.
For the purposes of this paper we only require terms linear in a weak
background supergravity field. Such terms have recently been found
in  \cite{Mark-Wati-4} for a system of many D0-branes in a weak
background field.  The result for D0-branes is derived
using a proposal for the linear terms in the general
background Matrix theory action motivated by the structure of the
linearized supergravity currents in Matrix theory
\cite{Dan-Wati-2,Mark-Wati-3}.

The results of \cite{Mark-Wati-4} can be carried over to the D3-brane
system by T-dualizing on a 3-torus and taking the limit of infinite
torus volume in the IIB theory.  For the case of the dilaton
field, the complete set of couplings is given by \cite{Mark-Wati-5}
\bea 
S_\phi &=& T_3
\int d^4 x \sum_{n=0}^{\infty}{1 \over n!} \{\partial_{l_1} \cdots
\partial_{l_n} \phi(x,0) \} \{{1 \over 6}T^{ii(l_1...l_n)} - {1 \over
3}T^{\hat{a}\hat{a}(l_1...l_n)} - {1 \over 3}T^{+-(l_1...l_n)}
\}\nonumber\\ &\equiv& T_3\int d^4 x \sum_{n=0}^{\infty} {1 \over n!}
\{\partial_{l_1} \cdots \partial_{l_n} \phi(x,0) \} A^{l_1 \cdots l_n}
\label{eq:matrix}
\eea 
where $T^{\mu \nu (l_1 \cdots l_n)}$ are T-dualized versions of
the Matrix theory expressions for the multipole moments of the $D=11$
DLCQ supergravity stress-energy tensor that were shown to appear
coupled to the background metric in the action for Matrix theory in a
general background. Here, the index $\hat{a}$ runs from 1 to 3 while
the remaining indices are $SO(6)$ indices running from 4 to
9. Explicit expressions for the $T$'s were determined in
\cite{Dan-Wati-2,Mark-Wati-4} by comparing the one-loop Matrix theory
interaction potential between two arbitrary objects with the tree
level supergravity result. Using those results, we may write
\footnote{ Here and throughout the rest of this work, indices
$a,b,\dots = 0,1,2,3$ are world-volume indices on the brane while
$i,j,\dots$ and $p,q,\dots$ are transverse indices running from 4 to
9. Also, $(i_1 \cdots i_n)$ and $[i_n \cdots i_n]$ denote averaged
symmetrization and antisymmetrization respectively.  All quantities
are to be interpreted as their dimensional reduction from $D=10$, so
for example $F_{ij} \equiv i[X^i,X^j]$ and $D_i \Theta \equiv i[X_i,
\Theta]$. }
\[ 
A^{l_1 \cdots l_n} = \str (\{{1 \over 4} F_{ab} F_{ab} - {1 \over 4} 
F_{ij} F_{ij} + {1 \over 4} \bar{\Theta} \Gamma^i D_i \Theta\}X^{l_1} 
\cdots X^{l_n}) + A^{l_1 \cdots l_n}_{f}
\]
Here $\str$ denotes an average of all possible orderings of the 
expressions $F$, $X$, $\Theta$ and $D \Theta$ in the trace. The terms 
$A^{l_1 
\cdots l_n}_{f}$ are a set of additional terms involving fermion 
fields which appear for $n>0$. For $n=1$, the explicit expression may be 
determined from the results of \cite{Mark-Wati-4} and is
\[
A^{l}_{f} = -{1 \over 16}\str (F_{ab} \bar{\Theta} \Gamma^{[abl]} \Theta 
- F_{ij} \bar{\Theta} \Gamma^{[ijl]} \Theta)
\]
The terms for $n>1$ could be determined by extending the Matrix theory
calculation in \cite{Mark-Wati-4} to higher orders in $1/r$, but we
will determine them more efficiently below from the purely bosonic
terms using supersymmetry and a connection to the AdS/CFT
correspondence.

    For a dilaton field $\phi(x^a,x^i)$, the $l$th partial wave is 
precisely the part whose expansion in transverse coordinates $x^i$ is in 
the $l$ index symmetric traceless representation of $SO(6)$. To isolate 
the terms in the action which couple to a particular dilaton partial wave 
we may rearrange the terms in (\ref{eq:matrix}) as
\beas
S_\phi & = &
T_3\int d^4 x\left.[ \{\phi(x,0)A + {1 \over 12} \{(\partial_i)^2 \phi(x,0)\}
A^{kk} + \dots \}\right.\\ 
&&\hspace{0.7in}
+ \{ \{\partial_i\phi(x,0)\} A^i + {1 \over 16} \{(\partial_i)^2
\partial_k \phi(x,0) \} \delta^{(ij}A^{k)ll} + \dots \}\\ 
&&\hspace{0.7in}
+ {1 \over 2} \{\partial_i \partial_j \phi(x,0)\} (A^{ij} - {1 \over 6} 
\delta^{ij} A^{kk}) + \cdots\\
&&\left. \hspace{0.7in}+ \cdots \right.]
\eeas
Here, the first line gives the coupling of the $s$-wave part of the
dilaton, the second line gives the coupling of the $l=1$ part, and so
forth. For each $l$, the leading low-energy cross section will come
only from the  terms with $l$ derivatives on $\phi$, since additional
derivatives on $\phi$ will result in additional powers of
$\omega$. Therefore, we define operators \footnote{Here, $C^{i_1
\cdots i_n}_{p_1 \cdots p_n}$, whose explicit form is given in the
appendix, is a combination of delta functions which picks off the
symmetric traceless part of any operator with $l$ $SO(6)$ indices.}  
\beas
{\cal O}^{k_1 \cdots k_n} &=& A^{k_1 \cdots k_n} - \{ {\rm traces} \}\\
&\equiv& A^{p_1 \cdots p_n} C^{k_1 \cdots k_n}_{p_1 \cdots p_n}
\eeas
such that the low-energy contribution to the $l$-wave absorption cross 
section will be determined by the term
\be
\label{eq:oper}
S_l = T_3 \int d^4 x {1 \over n!} \{ \partial_{k_1} \cdots \partial_{k_n} \phi 
(x,0)\} {\cal O}^{k_1 \cdots k_l}
\ee
To determine the remaining fermionic terms in the operators ${\cal
O}$, we now make a connection with the AdS/CFT correspondence for D3
branes.

   In the correspondence between large $N$ $D=4$, ${\cal N} = 4$ super 
Yang-Mills theory and type IIB supergravity on $AdS^5 \times S^5$ 
\cite{Maldacena-AdS,gkp-2,Witten-AdS1}, 
gauge theory operators corresponding to the complete spectrum of 
Kaluza-Klein modes of the supergravity fields have been found. These 
operators lie in short multiplets of the superconformal group and may be 
obtained by acting with various combinations of the $D=4$ supercharges 
$Q$ and $\bar{Q}$ (up to four of each) on the chiral primary operators 
\be
\label{eq:primary}
{\cal O}^{{\rm cp}}_n=
\tr (X^{p_1} \cdots X^{p_n}) C^{i_1 \cdots i_n}_{p_1 \cdots p_n} 
\ee   
In \cite{deAlwis,Das-Trivedi} it was conjectured that
the gauge theory operators coupling to the 
various supergravity modes may also be determined by expanding the 
Born-Infeld action for a D3 brane about the AdS background. In the case 
of the dilaton field, apart from some power of $r/R$, the operator 
determined in this way is exactly the same as the operator which couples 
to the dilaton in the Born-Infeld action expanded about flat space, 
since the dilaton does not mix with any other fields in either
picture. Hence, the operator we are interested in should be obtainable
by taking a supersymmetry variation on the chiral primary fields
above.  

More precisely, it may be seen from the analysis in
\cite{Gunaydin-Marcus} and \cite{krv} (the table in
\cite{Intriligator} is useful in relating the results in these papers
to the 4D theory) that the particle corresponding to the $l$th
partial wave of the dilaton couples to an operator obtained by
applying four supercharges of the same chirality to the primary
operator in (\ref{eq:primary}) with $n=l+2$. From the $D=10$ point of
view, both $Q$ and $\bar{Q}$ are contained in the Majorana-Weyl
supercharge $Q_\alpha$, so the operator coupling to the $l$th partial
wave of the dilaton field is contained in the operator \be
\label{eq:fourQ}
Q_\alpha Q_\beta Q_\gamma Q_\delta \tr (X^{p_1} \cdots X^{p_{l+2}}) 
C^{mni_1 \cdots i_{l}}_{p_1 \cdots p_{l+2}}
\ee
We use conventions in which the $D=10$ supersymmetry transformation
rules are\footnote{we include explicitly the projection operator $P =
(1+\Gamma^{11})/2$ in terms of which the Weyl condition is $P_{\alpha
\beta} Q_\beta = Q_\alpha$}
\beas
Q_\alpha A^\mu &=& i(\Gamma^0 \Gamma^\mu)_{\alpha \beta} \Theta_\beta\\
Q_\alpha \Theta_\beta &=& {i \over 2} (\Gamma^{[\mu \nu]} P)_{\beta 
\alpha} F_{\mu \nu}
\eeas

The operator we are interested in is a Lorentz scalar and a 
traceless $l$-index symmetric tensor of $SO(6)$, so our desired operator 
is actually obtained from the above expression by contracting the
extra indices with a combination of 10D gamma matrices of the form
\[
A^{mn}_{\alpha \beta \gamma \delta}.
\]
In principle, the linear combinations of terms in (\ref{eq:fourQ}) in
which we are interested can be determined from group theory, using the
results of \cite{Gunaydin-Marcus}.  These terms can also be isolated
by performing a component expansion of polynomials in superfields as
in \cite{flz}.
We find it easier in practice to
simply find a combination of gamma matrices which correctly reproduce
the bosonic and two-fermion terms described above.  This is achieved
by contracting (\ref{eq:fourQ}) with \footnote{The antisymmetrization in 
fermion indices  is required since we are trying to reproduce the action of 
four $D=4$ supercharges of like chirality, which anticommute. This 
restricts to antisymmetric matrices for which $\{\Gamma^{[\mu \nu 
\lambda]} \Gamma^0\}_{\alpha \beta}$ form a basis when sandwiched between 
Majorana Weyl spinors.}
\beas
A^{m n}_{\alpha \beta \gamma \delta} & =& 
{1 \over 3 \cdot 2^{10} \cdot 
(l+2)(l+1)} 
\left( \{\Gamma^{[abm]} 
\Gamma^0 \}_{[\alpha \beta} \{\Gamma^{[abn]} \Gamma^0 \}_{\gamma 
\delta]}- \{\Gamma^{[ijm]} \Gamma^0 \}_{[\alpha \beta} \{\Gamma^{[ijn]} 
\Gamma^0 \}_{\gamma \delta]}\right)
\eeas

This gives us the complete operator coupling to the the $l$th partial
wave of the 
dilaton, which may now be computed to be
\bea
\label{eq:fullop}
{\cal O}^{i_1 \cdots i_l} &= & \left\{ {1 \over 4} \str (\{ F_{ab} F_{ab} 
- F_{ij} F_{ij} +  \bar{\Theta} \Gamma^i D_i \Theta \} X^{p_1} 
\cdots X^{p_l}) \right. \\
&&\hspace{0.1in}
- { l \over 16} \str (\{F_{ab} \bar{\Theta} \Gamma^{[abp_1]} \Theta - 
F_{ij} \bar{\Theta} \Gamma^{[ijp_1]} \Theta \} X^{p_2} \cdots X^{p_l})
 \nonumber\\
&&\hspace{0.1in}
+ \left. { l(l-1) \over 768} \str (\{\bar{\Theta} \Gamma^{[abp_1]} 
\Theta
\bar{\Theta} \Gamma^{[abp_2]} \Theta - \bar{\Theta} \Gamma^{[ijp_1]}
\Theta \bar{\Theta} \Gamma^{[ijp_2]} \Theta \} X^{p_3} \cdots X^{p_l}) 
\right\} C^{i_1 \cdots i_l}_{p_1 \cdots p_l}
\nonumber 
\eea
Terms in the first line arise from the four supersymmetry generators
acting on either one or two of the $X$'s in (\ref{eq:primary}) and
appear for any $l$. In writing these terms, we have used the equations
of motion, written compactly using $D=10$ indices as 
\[
\Gamma^\mu D_\mu \Theta = 0, \; \; \; \; D_\mu F_{\mu \nu} = i \bar{\Theta} 
\Gamma^\nu \Theta
\]
to rewrite terms in a form with no world-volume derivatives acting on
$F$ or $\Theta$ (recall that $D_i \Theta \equiv i [X^i, \Theta]$). Terms in the 
second line result when the $Q$'s are spread over three separate $X$'s and 
appear for $l \ge 1$. Finally, the four fermion terms come when each 
supersymmetry generator acts on a different $X$ and therefore appear only for $l 
\ge 2$.  
   
For each of the dilaton partial waves, we have now determined the
complete form of the non-abelian operators ${\cal O}$ which determine the
leading term in the low-energy expansion of the absorption cross
section.

\section{World volume absorption}

In this section, we use the operator determined in the previous section 
to calculate the cross section for absorption of the $l$th partial wave 
of the dilaton field by a set of $N$ coincident parallel D3 branes. 

The most obvious way to proceed, and the method originally used in
\cite{Klebanov-absorption} to show agreement between the world volume
and supergravity approaches for the s-wave absorption, is to treat the
dilaton as a time dependent perturbation in the world-volume theory
and calculate the transition amplitude to each possible set of final
particles on the brane, summing over the various contributions in the
usual way to obtain a cross section. However, as explained in
\cite{Gubser-Klebanov}, it turns out that there is a simpler method
exploiting the fact that the cross section arising from a given
operator is simply related to the two-point function of that operator
on the brane. For a canonically normalized scalar coupling to the
brane through an interaction
\[
S = \int d^4 x \phi (x,0) {\cal O} (x)
\]
the precise relation is given by
\be
\label{eq:disc}
\sigma = \left. {1 \over 2 i \omega}  {\rm Disc}\; \Pi (p) \right|_{-p^2 = 
\omega^2 - i \epsilon}^{-p^2 = \omega^2 + i \epsilon}
\ee
Here, $\omega$ is the energy of the particle, and 
\[
\Pi(p) = \int d^4 x e^{i p \cdot x} \langle {\cal O}(x) {\cal O} (0) 
\rangle
\]
which depends only on $s=p^2$. To evaluate (\ref{eq:disc}) we extend 
$\Pi$ to  complex values of $s$ and compute the discontinuity of $\Pi$ 
across the real axis at $s=\omega^2$. This method has the advantage that 
it is not necessary to determine all of the distinct final particle 
states or sum over the polarizations, which would be rather complicated 
for large values of $N$ and $l$. 

We now use this method to calculate the absorption cross section for 
each partial wave of the dilaton field. We assume that the dilaton is 
normally incident on the brane in the $9$ direction so that
\[
\phi(x) = e^{i \omega (x^9-t)}
\]
From (\ref{eq:oper}), we see that the absorption cross section for the $l$th 
partial wave is determined by the two-point function of the operator 
\[
{\cal O}_l = T_3 {\omega^l \over l!} {\cal O}^{99 \cdots 9}
\]
{}From (\ref{eq:fullop}), we note that ${\cal O}_l$ has terms involving $l+2$ 
or more fields, so the leading contribution to $\langle {\cal O}_l(x) 
{\cal O}_l (0) \rangle$ will be an $l+1$ loop planar
diagram with each field in the operator at $x$ contracted with a field
in the operator at $0$. We can ignore all contributions from operators 
containing commutators $F_{ij}$ and $D_i \Theta$ since these contain more than 
$l+2$ fields and will come in at higher order in $(g_{YM}^2 N)$. The terms which 
do contribute are a bosonic term
\[
{\cal O}_l^{\rm bos} \equiv {T_3 \omega^l \over 4 l!} \str ( F_{ab} F_{ab}  
X^{p_1} \cdots X^{p_l}) C^{\vec{9}}_{p_1 \cdots p_l} \ ,
\]
a two fermion term
\[
{\cal O}_l^{2\Theta} \equiv -{ T_3 \omega^l \over 16 
(l-1)!}\str (F_{ab} \bar{\Theta} \Gamma^{[abp_1]} \Theta  X^{p_2} \cdots 
X^{p_l}) C^{\vec{9}}_{p_1 \cdots p_l}
\ , \]
and a four fermion term,
\[
{\cal O}_l^{4\Theta} \equiv  { T_3 \omega^l \over 768(l-2)!} 
\str (\{\bar{\Theta} \Gamma^{[abp_1]} \Theta \bar{\Theta} \Gamma^{[abp_2]} 
\Theta - \bar{\Theta} \Gamma^{[ijp_1]} \Theta \bar{\Theta} 
\Gamma^{[ijp_2]} \Theta \} X^{p_3} \cdots X^{p_l}) C^{\vec{9}}_{p_1 
\cdots p_l} \nonumber
\ .\]
The complete two point function is the sum of the two-point functions of 
each of these operators since there are no cross terms at leading order.  
\\ \\
{\bf Propagators}
\\ \\
To evaluate the two-point functions at leading order, all we need to 
know are the propagators of the various fields. In $D=10$ language, 
choosing a gauge fixing term which enforces the Feynman gauge, the 
quadratic action which determines the propagators is simply 
\[
S = T_3 \int d^4 x \tr (-{1 \over 2} A_b (\partial_a)^2 A_b - {1 \over 
2} X_i (\partial_a)^2 X_i - {1 \over 2} \bar{\Theta} \Gamma^a \partial_a 
\Theta)
\]
In terms of the scalar propagator
\[
\Delta(x-y) \equiv {1 \over 4 \pi^2 |x-y|^2}
\]
the propagators for the various fields are \footnote{In these 
expressions, the indices $k,l,m,n$ are $U(N)$ indices}
\beas
\langle X_i^{kl}(x) X_j^{mn}(y)\rangle &=& {1 \over T_3} \delta_{ij} 
\delta^{kn} \delta^{lm} \Delta(x-y)\\
\langle A_a^{kl}(x) A_b^{mn}(y)\rangle &=& {1 \over T_3} \delta_{ab} 
\delta^{kn} \delta^{lm} \Delta(x-y)\\
\langle \Theta_{\alpha}^{kl}(x) \Theta_{\beta}^{mn}(y)\rangle &=& {1 \over T_3} 
(P 
\Gamma^a \Gamma^0)_{\alpha \beta} \delta^{kn} \delta^{lm} \Delta(x-y)\\
\eeas
Note that the projection matrix $P$, defined above, appears in the 
fermion propagator, since half of the components of each spinor are 
zero. From the gauge field propagator, we also have to leading order in 
$1/x$ that
\[
\langle F_{ab}^{kl}(x) F_{cd}^{mn}(y)\rangle = {4 \over T_3} 
\delta^{kn} \delta^{lm} \partial_{[a}\delta_{b] [c} \partial_{d]}  
\Delta(x-y)\\
\] 
\\
{\bf Bosonic contribution}
\\ \\ 
We first compute the two-point function of the bosonic operator. We have 
\beas
\Pi_l^{\rm bos}(x) &=& \langle {\cal O}_l^{\rm bos}(x) {\cal O}_l^{\rm bos}(0) 
\rangle\\
&=& {T_3^2 \omega^{2l} \over 16 (l!)^2 } \; C_{\vec{p}}^{\vec{9}} 
C_{\vec{q}}^{\vec{9}} \; \langle \str (F_{ab} F_{ab} X^{p_1} \cdots 
X^{p_l})_x \str (F_{cd} F_{cd} X^{q_1} \cdots X^{q_l})_0 \rangle
\eeas
Note that since the $X$'s in each symmetrized trace contract with a 
totally symmetric tensor $C^{\vec{9}}_{\vec{p}} \equiv C^{9 \cdots 
9}_{p_1 \cdots p_l}$, we need only average over the $(l+1)$ orderings of 
operators in which one $F$ is fixed in the first position by cyclicity 
of the trace and the other runs over positions 2 through $l+2$.
By Wick's theorem, the correlator for each ordering of operators in the 
two symmetrized traces is evaluated by summing over all possible 
contractions matching the operators in the first trace to those in the 
second trace. However, only those contractions which match up the 
operators in reverse cyclic order contribute with the maximal power of 
$N$, namely $N^{l+2}$. For each of the $(l+1)$ orderings of operators in 
the first trace there will be exactly such 2 contractions with the sum 
of operators in the second trace.\footnote{These may come from two 
different contractions with the same operator for orderings such as 
$\tr (FXFX)$ or a single contraction with two different operators for 
orderings which are not invariant under a cyclic shift by $(l+2)/2$ 
positions.} All of these contributions are identical, so the 
symmetrizations result in a factor $2(l+1)/(l+1)^2=2/(l+1)$, and we have
\beas
\Pi_l^{\rm bos}(x) &=& {T_3^2 \omega^{2l} \over 16 (l!)^2 } 
\; C_{\vec{p}}^{\vec{9}} C_{\vec{q}}^{\vec{9}} \; {2 N^{l+2} \over l+1} \\ 
& & 
\hspace{0.2in} \times \langle F_{ab}(x)F_{cd}(0)\rangle \langle 
F_{ab}(x)F_{cd}(0)\rangle \langle X^{p_1}(x) X^{q_1}(0) \rangle \cdots 
\langle X^{p_l}(x) X^{q_l}(0) \rangle\\
&=& {T_3^{-l} \omega^{2l} N^{l + 2} \over l!(l+1)!} C_{p_1 \cdots 
p_9}^{\vec{9}} C_{p_1 \cdots p_9}^{\vec{9}}\Delta^l(x) \left( \partial_a 
\partial_b \Delta(x) \partial_a \partial_b \Delta(x) + {1 \over 2} 
\partial^2 \Delta(x) \partial^2 \Delta(x)\right)
\eeas
In the second line, we have already evaluated the contractions of $U(N)$ 
delta functions to give $N^{l+2}$, so the correlators there have the 
values of $U(1)$ correlators. In the last line, the term involving 
$\partial^2 \Delta(x) \propto \delta(x)$ will give a constant 
contribution to $\Pi(p)$ so we can ignore it for the purposes of 
computing the discontinuity. The evaluation of $C_{p_1 \cdots 
p_9}^{\vec{9}} C_{p_1 \cdots p_9}^{\vec{9}}$ is described in detail in the 
appendix. The simple result is that
\[
C_{p_1 \cdots p_l}^{\vec{9}} C_{p_1 \cdots p_l}^{\vec{9}} = {(l+2)(l+3) 
\over 3 \cdot 2^{l+1} }
\]
Thus, we find
\be
\label{eq:bos2pt} 
\Pi_l^{\rm bos}(x) = {\kappa^{l} \omega^{2l} N^{l+2} (l+2)(l+3) \over  
2^{3l+1}\pi^{5l/2+4}l!(l+1)! |x|^{2l+8}}
\ee
where we have substituted $T_3 = \sqrt{\pi}/\kappa$. 
Using the result (see, for example \cite{Gubser-Hashimoto}) that
\[ 
\left. {\rm Disc}\; \left( \int d^4 x {e^{i p \cdot x} \over |x|^{2m+4}} 
\right) \right|_{-p^2 = \omega^2 -i\epsilon}^{-p^2 \omega^2 +i\epsilon}  
= {2\pi^3i \omega^{2m} \over 4^m m! (m+1)!} 
\]
we may now use (\ref{eq:disc}) to give our final result for the cross 
section 
arising from ${\cal O}_l^{\rm bos}$ as \footnote{We include an extra factor 
of $2 \kappa^2$ relative to the formula (\ref{eq:disc}) since the 
dilaton field is not canonically normalized due to the factor of 
$1/2\kappa^2$ in front of the supergravity action.}
\bea
\sigma_l^{\rm bos} &=& {2\kappa^2 \over 2i \omega} \left. {\rm Disc}\; 
\Pi_l^{\rm bos} (p) \right|_{-p^2 = \omega^2 - i \epsilon}^{-p^2 = \omega^2 
+ i \epsilon}\nonumber\\
&=& {N^{l+2} \kappa^{l+2} \omega^{4l+3} \over 2^{5l+4} \pi^{5l/2+1} 
l!((l+1)!)^2(l+2)!}\nonumber\\
&=& {6 \over (l+2)(l+3)} \sigma^l_s
\label{eq:boscross}
\eea 
where $\sigma^l_s$ is the cross section
(\ref{eq:semiclassical-absorption}) computed from classical  
supergravity. Recalling that the two and four fermion operators only 
contribute for $l \ge 1$, we see that we have reproduced the agreement 
for the $l=0$ case originally found in \cite{Klebanov-absorption}. For 
$l>0$, where we expect 
additional contributions from the other operators, our result is safely 
less than $\sigma^l_s$, so we do not find the problem encountered in 
\cite{gkt}.  
\\ \\
{\bf Two-fermion contribution}
\\ \\
We now calculate the two-point function of the two fermion operator 
${\cal O}_l^{2 \Theta}$ to determine its contribution to the cross 
section. The calculation is similar to the bosonic two-point function so 
we will be brief. We have
\beas
\Pi_l^{2 \Theta}(x) &\equiv& \langle {\cal O}_l^{2 \Theta}(x){\cal 
O}_l^{2 \Theta}(0)\rangle\\
&=& {T_3^2 \omega^{2l} \over 16^2 ((l-1)!)^2 } \; C_{\vec{p}}^{\vec{9}} 
C_{\vec{q}}^{\vec{9}} \; (\Gamma^0 \Gamma^{[abp_1]})_{\alpha \beta} 
(\Gamma^0 \Gamma^{[cdq_1]})_{\gamma \delta}\\
& & \hspace{ 0.2in} \times \langle \str (F_{ab} \Theta_\alpha  \Theta_\beta 
X^{p_2} \cdots X^{p_l})_x \str (F_{cd} \Theta_\gamma \Theta_\delta X^{q_2} 
\cdots X^{q_l})_0 \rangle
\eeas
Here, for each of the $l(l+1)$ orderings in the first symmetrized trace, 
there are two terms in the second symmetrized trace (related by 
switching the $\Theta$'s) which have the correct ordering to give a 
non-vanishing set of contractions with the maximal power of $N$. Again, 
all contributions are identical, due to the symmetry of 
$C_{\vec{p}}^{\vec{9}}$ and the antisymmetry in the fermionic indices of 
$(\Gamma^0 \Gamma^{[abp_1]})_{\alpha \beta}$, so we get a factor 
$2/l(l+1)$ from the symmetrizations and find
\beas
\Pi_l^{2\Theta}(x) &=& {T_3^2 \omega^{2l} \over 16^2 ((l-1)!)^2 } 
\; C_{\vec{p}}^{\vec{9}} C_{\vec{q}}^{\vec{9}} \; {2 N^{l+2} \over l(l+1)} 
(\Gamma^0 \Gamma^{[abp_1]})_{\alpha \beta} (\Gamma^0 
\Gamma^{[cdq_1]})_{\gamma \delta}\\
 & & \hspace{0.2in} \times\langle F_{ab}(x)F_{cd}(0)\rangle \langle 
\Theta_\alpha (x) \Theta_\delta(0)\rangle \langle \Theta_\beta (x) 
\Theta_\gamma(0) \rangle \langle X^{p_2}(x) X^{q_2}(0) \rangle \cdots 
\langle X^{p_l}(x) X^{q_l}(0) \rangle\\
&=&{T_3^{-l} \omega^{2l} N^{l+2} \over 2^5 (l-1)!(l+1)! } C_{p_1 p_2 
\cdots p_l}^{\vec{9}} C_{q_1 p_2 \cdots p_l}^{\vec{9}} \Delta^{l-1}(x) 
\partial_{[a}\delta_{b] [c} \partial_{d]}  \Delta(x) \partial_e 
\Delta(x) \partial_f \Delta(x)\\
& & \hspace{0.2in}\times \tr ( \Gamma^{[abp_1]} P \Gamma^e \Gamma^{[cdq_1]} P 
\Gamma^f)
\eeas
The projection matrices in the trace\footnote{For multiple $P$'s in a
trace, as long as they are all  
separated by an even number of $\Gamma$ matrices, we may bring them 
together into a single $P$ since $P$ commutes with any pair of 
$\Gamma$'s and $P^2=P$. Then, since $P = (1 + \Gamma^{11})/2$ we will 
just get a factor of $1/2$ unless there are at least 10 other $\Gamma$'s 
with distinct indices.} just serve to reduce its value by 
1/2, and we may evaluate the trace over the 
remaining gamma matrices by the usual rules to find
\[
\tr ( \Gamma^{[abp_1]} P \Gamma^e \Gamma^{[cdq_1]} P \Gamma^f)  
\rightarrow  \delta_{p_1q_1}(128\delta_{df} \delta_{be} \delta_{ac} + 32 
\delta_{ef} \delta_{cb} \delta_{da})
\]
Note that the two sides of this expression are not equal, but equivalent 
when appearing in the two-point function above, since we have used the 
antisymmetry of the index pairs $[ab]$ and $[cd]$ and the symmetry of 
index pair $(ef)$ in order to simplify the trace. Inserting this trace 
into the expression above and simplifying, we find that 
\[
\Pi_l^{2 \Theta}(x) = l \cdot \Pi_l^{\rm bos}(x) 
\]
where $\Pi_l^{\rm bos}$ is the bosonic two-point function given in 
(\ref{eq:bos2pt}). We 
therefore see immediately from (\ref{eq:boscross}) that the contribution 
of the two-fermion operator to the cross section is 
\[
\sigma_{2 \Theta}^l = {6l \over (l+2)(l+3)}\sigma^l_s
\]
The total cross section so far,
\[
{6(l+1) \over (l+2)(l+3)}\sigma^l_s
\]
agrees with the classical result for $l=0$ and $l=1$, and is less than 
the supergravity result for $l \ge 2$, consistent with the fact that  
the four fermion operator is only present for $l \ge 2$.
\\ \\
{\bf Four fermion contribution}
\\ \\
Finally, we calculate the contribution to the cross section from the 
four fermion operator which appears for $l>1$. In this case, the 
operator ${\cal O}_l^{4\Theta}$ has two pieces, so we have to calculate 
the two-point function of each of the pieces as well as a cross term. 
These three correlators differ only in the indices on the $\Gamma$ 
matrices, so we may write them together as
\beas
\Pi_l^{4 \Theta}(x) &\equiv& \langle {\cal O}_l^{4 \Theta}(x){\cal 
O}_l^{4 \Theta}(0)\rangle\\
&=& {T_3^2 \omega^{2l} \over 9 \cdot 2^{16} ((l-2)!)^2 } 
C_{\vec{p}}^{\vec{9}} C_{\vec{q}}^{\vec{9}}\\
& & \hspace{0.1in} \times
 \langle \str (\Theta_\alpha  \Theta_\beta \Theta_\gamma 
\Theta_\delta X^{p_3} \cdots X^{p_l})_x \str (\Theta_{\hat{\alpha}}  
\Theta_{\hat{\beta}} \Theta_{\hat{\gamma}} \Theta_{\hat{\delta}} X^{q_3} 
\cdots X^{q_l})_0 \rangle\ \\
& & \hspace{0.1in} \times\left\{ (\Gamma^0 \Gamma^{[abp_1]})_{\alpha \beta} 
(\Gamma^0 \Gamma^{[abp_2]})_{\gamma \delta} (\Gamma^0 
\Gamma^{[cdq_1]})_{\hat{\alpha} \hat{\beta}} (\Gamma^0 
\Gamma^{[cdq_2]})_{\hat{\gamma} \hat{\delta}} \right.\\
& & \hspace{0.4in}+ (\Gamma^0 \Gamma^{[ijp_1]})_{\alpha \beta} (\Gamma^0 
\Gamma^{[ijp_2]})_{\gamma \delta} (\Gamma^0 
\Gamma^{[klq_1]})_{\hat{\alpha} \hat{\beta}} (\Gamma^0 
\Gamma^{[klq_2]})_{\hat{\gamma} \hat{\delta}}  \\
& & \hspace{0.4in} \left. - 2 (\Gamma^0 \Gamma^{[abp_1]})_{\alpha \beta} 
(\Gamma^0 \Gamma^{[abp_2]})_{\gamma \delta} (\Gamma^0 
\Gamma^{[ijq_1]})_{\hat{\alpha} \hat{\beta}} (\Gamma^0 
\Gamma^{[ijq_2]})_{\hat{\gamma} \hat{\delta}} \right\} \\
\eeas
This time, each of the $(l+1)l(l-1)$ orderings in the first symmetrized 
trace may couple in a single way to 24 different terms in the second 
trace (related by permuting the $\Theta$'s), but this time not all of 
the contributions are equivalent. When we contract the fermion 
propagators with the $\Gamma$ matrices above, 1/3 of the terms give two 
traces over $\Gamma$'s while the remaining 2/3 give a single trace. The 
result is
\beas
\Pi_l^{4 \Theta}(x) &=& {T_3^{-l} \omega^{2l} N^{l+2} \over 9 \cdot 2^{16} 
((l-2)!)^2 } C_{p_1 p_2 p_3 \cdots p_l}^{\vec{9}} C_{q_1 q_2 p_3 \cdots 
p_l}^{\vec{9}}\Delta^{l-2}(x) \partial_e \Delta(x) \partial_f \Delta(x) 
\partial_g \Delta(x) \partial_h \Delta(x)\\
& &\hspace{0.1in}
\times {8 \over (l+1)l(l-1)} \left\{ (\tr \tr^{aa} - 2\tr^{aa}) 
+ 
(\tr \tr^{ii} - 2\tr^{ii}) - 2(\tr \tr^{ai} - 2\tr^{ai}) \right\}^{p_1 p_2 
q_1 
q_2 efgh}
\eeas
Here, the traces are defined as
\beas
\tr \tr^{aa} &\equiv& \tr (P\Gamma^e \Gamma^{[abp_1]} P\Gamma^f 
\Gamma^{[cdq_1]}) \tr (P \Gamma^g \Gamma^{[abp_2]} P \Gamma^h 
\Gamma^{[cdq_2]})   \\ & &\rightarrow   3\cdot 2^{11} \delta_{p_1 q_1} 
\delta_{p_2 q_2} \delta_{ef} \delta_{gh}\\
\tr^{aa} &\equiv& \tr (P\Gamma^e \Gamma^{[abp_1]} P\Gamma^f 
\Gamma^{[cdq_1]}P \Gamma^g \Gamma^{[abp_2]} P \Gamma^h \Gamma^{[cdq_2]}) 
  \\ & &\rightarrow   -3\cdot 2^9 \delta_{p_1 q_1} \delta_{p_2 q_2} 
\delta_{ef} \delta_{gh}\\
\tr \tr^{ii} &\equiv& \tr (P\Gamma^e \Gamma^{[ijp_1]} P\Gamma^f 
\Gamma^{[klq_1]}) \tr (P \Gamma^g \Gamma^{[ijp_2]} P \Gamma^h 
\Gamma^{[klq_2]})   \\ & &\rightarrow   3\cdot 2^{11} \delta_{p_1 q_1} 
\delta_{p_2 q_2} \delta_{ef} \delta_{gh}\\
\tr^{ii} &\equiv& \tr (P\Gamma^e \Gamma^{[ijp_1]} P\Gamma^f 
\Gamma^{[klq_1]}P \Gamma^g \Gamma^{[ijp_2]} P \Gamma^h \Gamma^{[klq_2]}) 
  \\ & &\rightarrow   - 3\cdot 2^9 \delta_{p_1 q_1} \delta_{p_2 q_2} 
\delta_{ef} \delta_{gh}\\
\tr \tr^{ai} &\equiv& \tr (P\Gamma^e \Gamma^{[abp_1]} P\Gamma^f 
\Gamma^{[ijq_1]}) \tr (P \Gamma^g \Gamma^{[abp_2]} P \Gamma^h 
\Gamma^{[ijq_2]})   \\ & &\rightarrow   0\\
\tr^{ai} &\equiv& \tr (P\Gamma^e \Gamma^{[abp_1]} P\Gamma^f 
\Gamma^{[ijq_1]}P \Gamma^g \Gamma^{[abp_2]} P \Gamma^h \Gamma^{[ijq_2]}) 
  \\ & &\rightarrow   9\cdot 2^9 \delta_{p_1 q_1} \delta_{p_2 q_2} 
\delta_{ef} \delta_{gh}\\
\eeas
Again, the evaluation of the traces is simplified using the fact that 
the sets of indices $(efgh)$, $(p_1 p_2)$, and $(q_1 q_2)$ are symmetric 
in the expression to which the traces are contracted. The rest of the 
evaluation is straightforward and in terms of the bosonic two-point 
function, we find
\[
\Pi_l^{4 \Theta}(x) = {l(l-1) \over 6} \cdot \Pi_l^{\rm bos}(x) 
\]
{}From (\ref{eq:boscross}), we may immediately read off the final 
contribution to the 
cross section to be 
\[
\sigma_{4 \Theta}^l = {l(l-1) \over (l+2)(l+3)}\sigma^l_s
\]
Combining the bosonic, two-fermion, and four-fermion contributions, we 
find the total cross section from the world volume calculation to be
\beas
\sigma^l_{tot} &=& \sigma_{\rm bos}^l + \sigma_{2 \Theta}^l + \sigma_{4 
\Theta}^l\\
&=&{6 \over (l+2)(l+3)}\sigma^l_s + {6l \over (l+2)(l+3)}\sigma^l_s + 
{l(l-1) \over (l+2)(l+3)}\sigma^l_s\\
&=&\sigma^l_s
\eeas
Thus, for all values of $l$, the total low-energy cross section for 
absorption of the $l$th partial wave of the dilaton by $N$ coincident 
D3-branes is exactly the same when computed in the world volume theory 
as when computed in classical supergravity.

\section{Conclusions}

In this paper we used the world-volume theory of many parallel
D3-branes to exactly reproduce the semiclassical absorption
cross section of an arbitrary higher partial wave of the dilaton
field.  This is the first time that such a correspondence has been
made precise for the absorption of higher partial waves by any D-brane
black hole configuration.  This result provides additional evidence
for the conjectured exact correspondence between the world-volume
theory of $N$ D3-branes and type IIB string theory on $AdS_5 \times
S^5$ \cite{Maldacena-AdS,gkp-2,Witten-AdS1}.  The fact that arbitrary
partial waves on the sphere $S^5$ are accurately described in the
D-brane gauge theory suggests a number of interesting directions for
further research.  In particular, these results indicate that incoming
wave packets of the supergravity fields can be localized on the
sphere in the asymptotic regime.  It would be interesting to study  in
more detail the
behavior of such localized wave packets in the D-brane gauge theory.

In performing the calculation in this paper, it was necessary to have
an exact formulation of the coupling of the D-brane world-volume
fields to the background supergravity fields.  We were able to
precisely fix the operators on the D3-brane world-volume which couple
linearly to derivatives of the background dilaton field by utilizing
recent results for similar operators in M(atrix) theory and the related
D0-brane theory in type IIA.  It has been suggested in various
contexts that the AdS/CFT correspondence and the M-theory/Matrix
correspondence are in some sense equivalent
\cite{Hyun,deAlwis-correspondence,Silva,Chepelev-are, 
Polchinski-M-theory,Townsend-matrix}.
The fact that similar operator structures appear coupling to
background fields in the two theories may help to make this
relationship more precise.  Certainly, the symmetrized trace structure
which appears in the supergravity operators found in
\cite{Mark-Wati,Dan-Wati-2,Mark-Wati-4} plays a key combinatorial role
in exact calculations in both theories, as seen in \cite{Mark-Wati-5}
and the present paper.  A fruitful direction for further progress may
be to use results from one of these correspondences in deriving new
information about the other, as we have here used Matrix theory
results to obtain new information in the D3-brane context.

The exact correspondence between the semiclassical gravity
calculation, which we expect to be valid for large $\gym N$, and the
super Yang-Mills calculation, which is an expansion to leading order
in $\gym N$, indicates that there is a non-renormalization theorem for
the two-point functions of all the operators ${\cal O}_l$ coupling to
$l$th partial waves of the dilaton.  Such a non-renormalization theorem
was proven in \cite{Gubser-Klebanov} for the two-point function of the
stress tensor, and in \cite{afgj} for the two-point function of the
R-symmetry current.  These operators lie in the same $p
= 2$ representation of the superconformal algebra $SU(2, 2 | 4)$ as
the operator ${\cal O}_0$ corresponding to the s-wave of the dilaton.
{}From general arguments based on supersymmetry \cite{Howe-West} it is
believed that all two-point functions of operators in this
representation are related by supersymmetry so that the
non-renormalization of the s-wave absorption amplitude is implied by
the non-renormalization theorems proven in
\cite{Gubser-Klebanov,afgj}.  For the operators coupling to the higher
partial waves there is as yet no analogous non-renormalization
theorem, although it is widely believed that all two and three-point
functions of operators in short representations of the superconformal
algebra are protected by non-renormalization theorems.  
Evidence for such non-renormalization theorems
was given in  \cite{lmrs}, where it was shown that the free field
calculation of the 3-point functions of the chiral primary operators
${\cal O}^{{\rm cp}}_p$ in (\ref{eq:primary}) 
agrees with the predictions of supergravity through the AdS/CFT
correspondence.  This calculation was somewhat different in spirit from
ours, though, because in \cite{lmrs} the overall normalization
of operators was left undetermined and only appropriate ratios
of correlators were shown to agree 
between weak and strong coupling. The advantage of using the absorption
cross sections to calculate two-point functions is that
the overall normalization of operators is completely fixed by 
comparing the coupling
of the throat region of the threebrane geometry to the bulk region and the 
corresponding coupling of the D3-branes to the bulk fields.

Perturbative evidence for the non-renormalization theorems
was given in
\cite{dfs2}, where it was shown that the first
perturbative correction to the two- and three-point functions of all the chiral
primaries vanishes
for all $p$.  We expect similar results to hold for the
descendant operators that we have constructed. In such calculations 
it will be 
important to use the complete vertex operators (\ref{eq:fullop}), 
including the parts that contain more than $2+l$ fields. 
Our results provide a strong piece of evidence for
the existence of non-renormalization theorems for two-point
functions; it would be very nice, however, to have a more direct
demonstration of these theorems and a better understanding of why they
occur.

The existence of an infinite family of non-renormalization theorems
for the two-point functions of short operators in the 4D super
Yang-Mills theory seems to be related to a similar infinite family of
non-renormalization theorems in the one-dimensional matrix quantum
mechanics theory underlying Matrix theory.  One piece of evidence
for the conjecture that Matrix theory describes light-front
supergravity is the agreement between the leading $v^4/r^7$ term in
the 1D super Yang-Mills effective potential describing the interaction
between a pair of D0-branes and the long-range supergravity effective
potential between a pair of gravitons with longitudinal momentum
\cite{DKPS,BFSS}.  This agreement arises due to a non-renormalization
theorem in the matrix quantum mechanics theory
\cite{pss,Lowe-constraints}.  In \cite{Dan-Wati-2,Mark-Wati-4} it is
shown that all linearized supergravity interactions are correctly
reproduced by one-loop terms in the matrix quantum mechanics theory,
suggesting an infinite family of non-renormalization theorems for
terms of the form $F^4 X^l/r^{7 + l}$.  It seems likely that similar
non-renormalization theorems occur in the effective action of the 4D
${\cal N} = 4$ gauge theory, generalizing the non-renormalization
theorem proven in \cite{Dine-Seiberg,Lowe-vu} for $F^4$ terms.
It is unlikely, though, that there are similar theorems for operators 
involving higher powers of $F$ than $F^4$ because such operators
do not belong to short multiplets (in the AdS/CFT correspondence
such operators are assumed to couple to massive string modes).
Note, however, that at least for the $SU(2)$ Matrix theory there appears to
be a non-renormalization theorem for the $v^6$ terms \cite{bbpt,pss2}.

Just
as non-renormalization of two-point functions in the 4D gauge theory
seems to correspond with the non-renormalization theorems in matrix
theory associated with linearized gravity interactions, there is
evidence for non-renormalization of three-point functions in the gauge
theory \cite{fmmr,Liu-Tseytlin,lmrs,dfs2,gkp-4} as well as for 3-body
interactions in Matrix theory \cite{pss2,Okawa-Yoneya}. 
A possible explanation for the non-renormalization of 2-point and
3-point functions in the $D = 4$ theory is given in \cite{Intriligator-Skiba}.
In both the AdS/CFT and matrix theory contexts, however, it appears
that there are no non-renormalization theorems for four-point
interactions.
In the AdS/CFT correspondence
there are corrections to supergravity 4-point functions
coming from explicit $O(\alpha'^3)$ corrections present in the string action
\cite{Banks-Green,Brodie-Gutperle,Intriligator-Skiba}.  
Analogous logarithmic corrections have been found in the 4-point
functions  of the super Yang-Mills theory \cite{dfmmr}.
Similarly, it does not seem to be possible to
extend the existing Matrix theory non-renormalization theorems to
4-body interactions \cite{deg2,Sethi-Stern-2}.  In the case of
Matrix theory, the absence of such non-renormalization theorems at
higher order would imply that agreement between matrix quantum
mechanics and supergravity might only be achieved through subtleties
in the large $N$ limit.  The suspected non-renormalization theorems
for 2- and 3-point interactions, which are as yet poorly understood,
form another interesting point of contact between Matrix theory and
the AdS/CFT correspondence.  It may be possible to use the
correspondence we have exploited in this paper between operators in
the two theories to achieve a better understanding of the structure in
supersymmetric gauge theory responsible for these non-renormalization
theorems.

In this paper we have only considered the leading term in the
absorption cross section for each partial wave $l$.  It would be very
interesting to study whether any of the higher order terms in the
semiclassical absorption result (\ref{eq:exact-l}) can be reproduced
from the D3-brane gauge theory.  It was suggested in \cite{ghkk} that
the nonabelian Born-Infeld action might give rise to this entire
series of terms.  It is argued in \cite{Gubser-Hashimoto}, however,
that string corrections to the NBI action will be needed to make the
correspondence precise beyond leading order.  This would not be too
surprising, as we have no reason to believe that the subleading
operators coupling to the higher partial waves of the dilaton will not
be renormalized.  In the past we have encountered many surprising
agreements, however, so it would be very interesting to extend the
analysis to subleading terms and to see if any further structure of
the absorption cross section can be understood on the D3-brane side.

\appendix

\section{Properties of the symmetric traceless tensor}

The symmetric traceless tensor $C^{i_1 \cdots i_n}_{p_1 \cdots p_n}$ is 
a combination of delta functions which projects onto the symmetric 
traceless part of any object with $l$ $SO(6)$ indices. For example,
\[
\tr (X^pX^qX^l)C^{ijk}_{pql} = \str (X^iX^jX^k - {3 \over 8} 
\delta^{(ij}X^{k)} X^lX^l)
\]
The precise definition of $C$ is
\be
\label{eq:cdef}
C^{i_1 \cdots i_n}_{p_1 \cdots p_n} =  \sum_{k=0}^{\lfloor n/2 \rfloor} 
a^n_k \; \delta_{(p_1 p_2} \cdots \delta_{p_{2k-1} p_{2k}} 
\delta_{p_{2k+1}}^{(i_n} \cdots \delta_{p_n)}^{i_{2k+1}} 
\delta^{i_{2k}i_{2k+1}} \cdots \delta^{i_2 i_1)}
\ee
By tracing over $i_1$ and $i_2$ and requiring that the result vanishes, 
we obtain a recursion relation for the coefficients $a^n_k$ which may be 
solved to give
\[
a^n_k = (- {1 \over 4})^k {n-k+1 \choose k+1} {k+1 \over n+1}
\]
\\
To evaluate the two-point functions in section 3, it is necessary to 
determine the value of
\[
c_n \equiv C^{9 \cdots 9}_{p_1 \cdots p_n} C^{9 \cdots 9}_{p_1 \cdots 
p_n} 
\]
We do this by noting that on the unit 5-sphere, we have
\be
\label{eq:gegen}
C^{9 \cdots 9}_{p_1 \cdots p_n} x^{p_1} \dots x^{p_n} = {1 \over 2^n 
(n+1)} C^2_n(x^9)
\ee
where $C^{\lambda}_n$ are the Gegenbauer polynomials, defined by
\[
\sum_{n=0}^{\infty} C^{\lambda}_n(x) \alpha^n = (1 - 2x\alpha + 
\alpha^2)^{-\lambda}
\] 
which play the same role for $S^{2 \lambda + 1}$ as the Legendre 
polynomials play for $S^2$. That is, they are the subset of spherical 
harmonics which arise in the the expansion of a function of only a single 
coordinate  (in our case, the plane wave $e^{i \omega x^9}$). The 
Gegenbauer polynomials thus obey an orthogonality relation, given by
\be
\label{eq:ortho}
\int_{-1}^1 dx C^{\lambda}_m(x) C^{\lambda}_n(x) (1-x)^{\lambda - 1/2}  = 
\delta_{mn} \; {\pi 2^{1-2 \lambda} \Gamma(n+2 \lambda) \over n! ( \lambda 
+n) (\Gamma(\lambda))^2}
\ee
To determine $c_n$, we square both sides of (\ref{eq:gegen}) and integrate 
over the unit five-sphere, evaluating the right side using the 
orthogonality relation (\ref{eq:ortho}), and the left side using the 
relation \cite{lmrs}
\[
\int_{S^5} x^{i_1} \cdots x^{i_{2m}} = {\omega_5 \over 2^{m-1} (m+2)!} 
\times \{ {\rm Sum \; of \; all \; index \; contractions} \}
\]
where $\omega_5$ is the area of a unit 5-sphere. The result is 
\beas
{\omega_5 n! \over 2^{n-1} (n+2)!} C^{9 \cdots 9}_{p_1 \cdots p_n} C^{9 
\cdots 9}_{p_1 \cdots p_n} &=& \left({1 \over 2^n (n+1)} \right)^2 
\omega_4 \int_{-1}^1 dx (1-x^2)^{3/2}C^2_n(x) C^2_n(x)\\
&=&\left({1 \over 2^n (n+1)} \right)^2 \omega_4 { \pi (n+3)! \over 8 n! 
(n+2) }
\eeas
and using $\omega_4 = 8\pi^2/3$, $\omega_5 = \pi^3$ our final result is 
\be
\label{eq:final}
\sum_{p_i = 4}^9 C^{9 \cdots 9}_{p_1 \cdots p_n} C^{9 \cdots 9}_{p_1 
\cdots p_n} = {(n+3)(n+2) \over 3 \cdot 2^{n+1}}
\ee
This expression is used in each of the two-point function evaluations to 
obtain a closed form for the final cross section.

\section*{Acknowledgments}

We would like to thank Steve Gubser and Samir Mathur for helpful
conversations.  WT would also like to thank the students in the MIT
special topics class 8.872, spring 1999. IRK is grateful to
the Physics Department of the Universities of Torino and Milano for hospitality 
during the final stages of this project. The work of IRK is supported
in part by the NSF grant PHY-9802484 and in part by the James
S. McDonnell Foundation Grant No. 91-48.  The work of WT is supported
in part by the A.\ P.\ Sloan Foundation and in part by the DOE through
contract \#DE-FC02-94ER40818.  The work of MVR is supported in part by
the Natural Sciences and Engineering Research Council of Canada
(NSERC).

%%%%%%%%%%%%%%%%%%%%%%

\bibliographystyle{plain}

\end{document}